# Commissioning strategies and methods

*John Galambos*
Spallation Neutron Source, Oak Ridge National Laboratory, USA

**Abstract**
Accelerator beam commissioning is a challenging and exciting period. It is generally the first integrated operation of the many systems in an accelerator and, most importantly, of the beam. First, general preparation is discussed. Then general methods for initial beam commissioning are described, including methods for transverse and longitudinal beam set-up. The particular emphasis here is on tuning methods for linear accelerators.

## 1 Introduction

Beam commissioning is one of the most exciting times in the accelerator life cycle—the birth of the beam. As with the delivery of a child, things can be not only exciting but also stressful. In order for the commissioning experience to go as smoothly as possible, proper preparation is important, along with following certain guiding principles. Tools and principles for commissioning are discussed here. In keeping with the thrust of the other lectures in this school, emphasis is given to the commissioning of high-power proton linear accelerators, but many of the concepts discussed here are general.

High-power, high-intensity accelerators tend to be large, expensive facilities. Often these construction projects tend to run behind the anticipated schedule, with added pressure and reduced time available for beam commissioning. Nevertheless, several recently completed facilities have had successful beam commissioning despite minimal available time. These include the Spallation Neutron Source (SNS) [1–3], the Japan Proton Accelerator Research Complex (J-PARC) [4], and more recently the Large Hadron Collider [5].

Beam commissioning refers to the initial transport and tuning of the beam through an accelerator. However, much of the effort to accomplish this is put in beforehand. The preparation activities will be discussed first. Next, the methods used to ensure proper magnet and RF set-up are reviewed. Finally, some overall commissioning strategies are discussed.

## 2 Commissioning preparations

In general, it is much easier to shake out software and equipment before the actual commissioning, in a controlled environment, rather than after commissioning starts. Having the eyes of a commissioning team looking at you while you try to mend a system that is holding everyone up is a situation to be avoided. Generally, system engineers will test their individual components with acceptance tests, but additional verification is useful. For example, magnets may appear to be working properly, but there are ample opportunities for polarity swaps. A useful check is to carry a Hall probe along the beam line with the magnets powered and verify that the quadrupoles and dipoles all have the correct polarity.

One important commissioning tool is an 'on-line' model that can be used in predictive ways to help understand and correct the state of the machine. One of the primary inputs to such a beam model is the field strength of the magnets in the accelerator. Engineers usually provide magnet control via prescribed-current control, but the physicist needs to read or prescribe the magnetic field strength or focusing strength. A careful magnetic-field–current mapping should be done for the magnets prior to commissioning, and the magnetic-field 'physics units' must made available to the model in real time.

When measuring the magnetic-field–current mapping in a magnet-measuring set-up before magnet installation, one can also perform hysteresis tests on magnets containing iron, to determine whether a magnet-cycling procedure is needed for reliable magnetic-field setting (e.g. one can determine how fast the cycling can be performed).

In addition to providing a physically meaningful interface to the magnets, providing a physically meaningful interface to the RF accelerating structures is also useful. This, however, is more difficult. For the RF structures, there are two important parameters: the phase with respect to the beam, and the amplitude. The precise setting of these parameters using beam-based methods will be discussed below, but a rough calibration of the RF amplitude can be made a priori, using RF power measurements and knowledge of the cavity shunt impedance [6].

Beyond equipment checks, software reliability is always a concern. Recently, 'virtual accelerators' have been prepared before beam commissioning to simulate the behaviour of the beam [7, 8]. These virtual accelerators are model-based representations of the beam. They receive input for magnet and RF settings through the accelerator control system, initialize a model appropriately, and perform a beam simulation calculation. Typically, the models are simple linear matrix representations of the beam. Space charge is often treated as a defocusing (linear) correction term. But an important metric is computational speed; in the control room, one cannot usually afford to wait minutes for feedback from a model. From the model output, the responses of the beam-line instruments (e.g. beam position monitor signals) are then generated and sent out through the control system. These virtual accelerator set-ups certainly cannot catch many equipment-related issues, but are useful for debugging high-level applications used to analyse and control beam behaviour.

Finally, beam simulations are a critical part of the preparation for commissioning. Most beam simulations during the accelerator design stage tend to focus on the final, high-power, operational scenarios. But understanding the behaviour of the beam during the initial low-intensity commissioning stage is also important. These simulations form the basis of the beam-commissioning plan, and are used to define the requirements for beam diagnostics. Having a notion of what to expect for off-normal settings is useful, as this will likely be the case during the initial set-up. For example, low-energy proton beams are easily debunched, and if an RF structure initially has a synchronous phase that debunches the beam, downstream strip-line pickups may not be excited. There may be nothing wrong with the pickup, even though there is no signal coming from it. The beam distribution for an off-normal RF setting is simply different from the design expectation for which the pickup may have been designed.

Most beam-commissioning methods involve varying an external input to the beam (e.g. a magnetic field, RF phase, or RF amplitude) and observing the downstream effect on the beam with an instrument. Having a priori simulations of what to expect is important. Many of the commissioning techniques discussed below involve comparing measured beam changes with calculations of expected changes, using parametric variations in the changes applied. By comparing the observed changes with model calculations, one can calibrate the control of the external input and set it to a design value.

## 3  Transverse beam set-up

### 3.1  Trajectory correction

One the first beam-commissioning tasks is sending the beam down the centre of a section of beam pipe. Trajectory correction (or orbit correction in a circular machine) is one of the most common types of tuning. It involves (1) generating a response matrix for the change in the beam position at a measurement point (i.e. the position of a Beam Position Monitor, BPM) with the strength of each dipole corrector element to be used, (2) measuring the beam position at each BPM, (3) solving for the corrector strengths required to best return the beam to the quadrupole centres (or other desired positions) along the beam line, and (4) applying the new corrector magnet settings and seeing if they work.

The response matrix can be generated in two ways: model-based or empirically. The advantage of the model-based approach is that it is fast. The model must be initialized with the magnet strengths in the beam line (correctors, focusing and bending magnets, etc.), and with the RF cavity set-up. Note that the latter requires knowledge of the synchronous phase of the beam relative to the RF (as RF cavities apply a transverse kick to off-axis proton beams), which is generally not known at the start of commissioning. So one can turn RF cavities off until they are properly set up, when doing initial trajectory correction, and then repeat the operation after the RF is set up. A disadvantage of the model-based approach is that it does not account for any BPM or quadrupole misalignments (at least initially—they can be added in to the model later). The empirical approach to response matrix generation has the disadvantage of being slower: one must apply a kick with each corrector element individually and measure the response of all downstream BPMs (or all BPMs in the case of a circular orbit). It also requires applying a large enough kick to generate a wave with a displacement sufficiently greater than the noise level of the BPMs (typically, a wave of displacements of a few millimetres is required to generate a clean response matrix), and this can sometimes be a high-loss exercise. The empirical approach has the advantage of including the effects of any quadrupole misalignments. Also, it accounts for any possible polarity swaps of dipole correctors and BPMs (if one is trying to centre the beam trajectory).

Regarding finding the solution for the best dipole corrector strengths in step 3 above, there are many approaches, ranging from simple least squares minimization of the beam displacements to sophisticated constrained optimization methods. The advantage of constrained optimization techniques is the straightforward inclusion of the limitations of the dipole corrector power supplies. With a sparse corrector deployment, the ability to minimize the beam trajectory displacement will likely be limited by the corrector power supplies. Another consideration in the solution is the weighting between the benefit gained in reducing the trajectory displacement versus the corrector strength applied. Sometimes only a small additional benefit in trajectory correction is gained from a large increase in a corrector power supply setting. These effects can be handled by appropriate weighting of the figure of merit used in the minimization procedure.

### 3.2 Orbit difference

During commissioning, many issues can cloud the beam-centring procedure, including polarity swaps of dipole correctors and BPMs. Trajectory difference techniques (or orbit difference techniques in the case of a circular device) provide a powerful technique for identifying these issues. The method is simple: apply a kick to the beam and compare the measured change in the beam trajectory with a model-predicted change. Although it is difficult to know the initial absolute position and angle of the beam at the start of a beam-line section, the prediction of the change in the trajectory is independent of the initial conditions of the beam (for a linear transport system). The concept is shown schematically in Fig. 1. The red line in this figure indicates the difference between the original beam trajectory (black) and the perturbed trajectory (blue). Note that the difference is zero upstream of the kick (for a non-ring application).

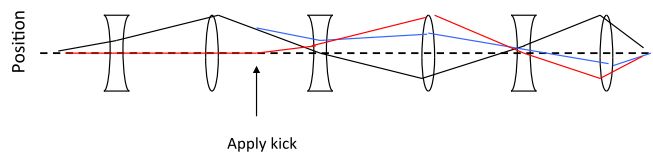

**Fig. 1:** Schematic illustration of the application of the trajectory difference technique. The original trajectory is shown in black, the trajectory after a kick has been applied is shown in blue, and the difference between the original and kicked trajectories is shown in red.

Figure 2 shows an example of an application of the trajectory difference technique. In this case one measured change at a BPM is opposite to the predicted change—a likely suspect for a BPM polarity issue. If, instead, all the measured beam position changes are opposite in sign to the expected changes, and only one corrector element shows this behaviour, there is likely to be a corrector polarity issue. It is also possible that the model-predicted and measured differences agree with each other, but both are incorrect. To determine if this is the case, one can vary the beam position at the location of an insertable device (such as a retractable-wire profile measurement device) and compare the measured sign change of the beam position at the insertable device with the expected sign change. This technique is useful for addressing systematic sign issues.

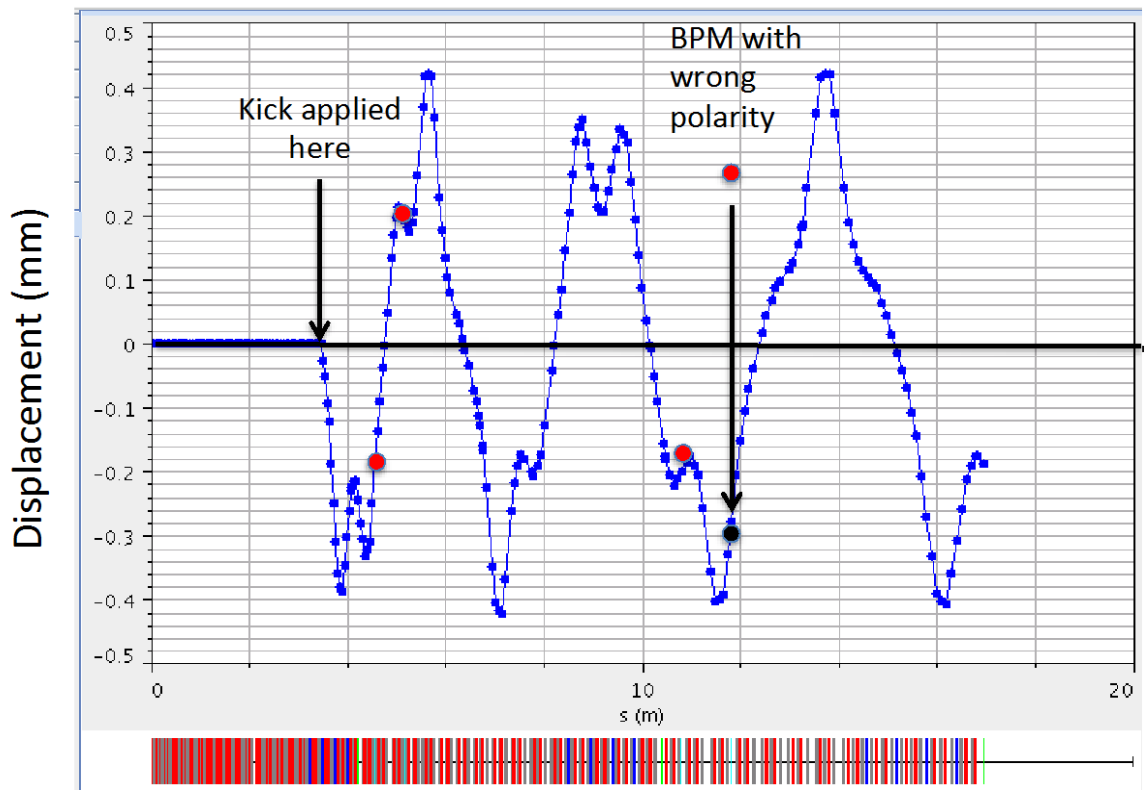

**Fig. 2:** Example of a trajectory difference. The line shows a model-predicted response of the trajectory to a dipole kick, and the large points show the measured BPM responses.

## 3.3 Transverse matching

Understanding the transverse beam size throughout a beam line and correcting it to the desired design lattice is a fundamental tuning process. A transverse beam mismatch increases the maximum beam size in the beam line, resulting in a closer approach to the aperture. Transverse matching optimizes the effective use of the aperture with respect to the core beam, and is a good starting step for beam loss reduction. Also, beam mismatch can cause halo growth [9].

Matching techniques require beam size measurements and independently adjustable quadrupoles. Typically, profile measurement stations are situated at the start of lattice transitions, and adjustable quadrupoles (often referred to as matching quads) are provided upstream of the beam measurement stations to facilitate matching. There are two primary approaches to transverse matching: (1) model-based methods and (2) a beam response matrix method.

## 3.3.1 Beam size measurement

For any transverse-matching algorithm, beam sizes along a lattice sequence must be measured. The most common method is to use wire scanner profile measurements [10, 11], which typically provide profiles of the beam distribution in the horizontal and vertical directions. Characteristic beam sizes must be obtained from these profiles, with the most common techniques being (1) calculation of the r.m.s. deviation from the centre of the beam, and (2) fitting the beam distribution with a Gaussian profile. The former technique is more general, as beam distributions are not always Gaussian. However, it is not easy to produce a robust generalization of this technique, and the results tend to be sensitive to the noise floor cut-off of the data. Usually, the Gaussian approximation works well. Figure 3 shows example linac beam profiles: one case is well approximated by a Gaussian profile but the other has a non-Gaussian beam halo. Additional pitfalls in the interpretation of profile data for use in transverse matching are discussed in Ref. [12].

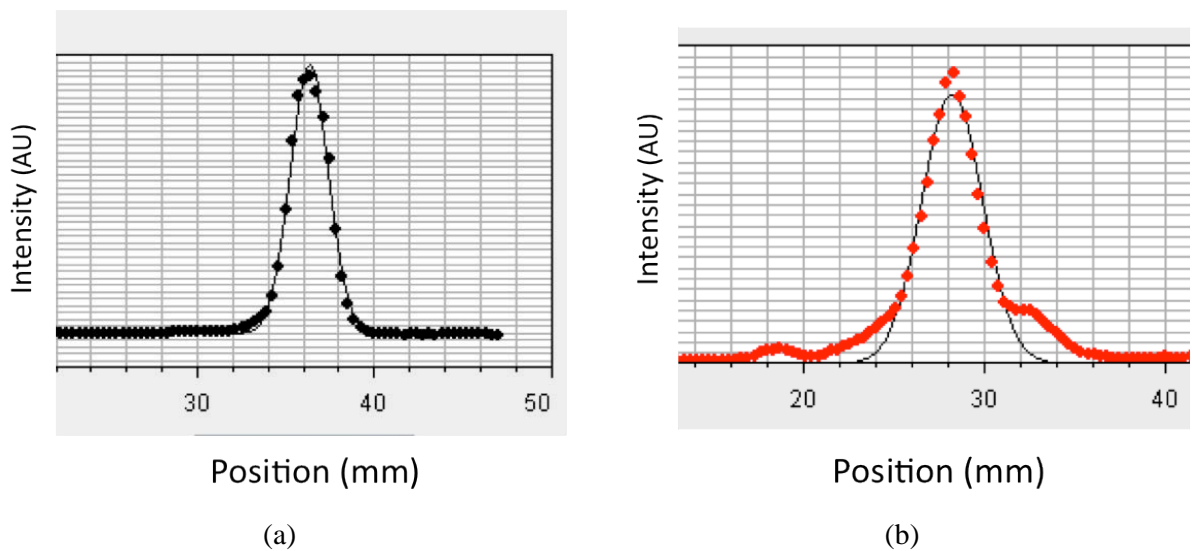

(a)                                    (b)

**Fig. 3:** Example of measurements of beam size from profile data. The dots show the wire-scanner-measured intensity, and the black lines show the best-fit Gaussian profile. Case (a) is an example at the beginning of the SNS linac (drift-tube linac section) where the beam is well represented by a Gaussian profile, and case (b) is an example at the end of the SNS linac where the beam has tails and is not well fitted by a Gaussian profile.

## 3.3.2 Model-based matching

An envelope model, given an initial set of Twiss parameters, can predict the r.m.s. beam size throughout a beam-line section. Envelope models (e.g. [13]) provide solutions for linear transport, often with corrections for space charge effects, which are important for high-intensity, low-energy beams. Given a set of beam size measurements in a beam line and knowledge of the lattice focusing strengths at the time of the measurements, one can solve for the initial Twiss parameters $\alpha$, $\beta$, and $\varepsilon$ [14] upstream of the measurements, so that the model-predicted beam sizes at the measurement points match the measured sizes best. The lattice region between the upstream Twiss solution point and the profile measurements should contain quadrupoles to be used for subsequent matching. At least three separate beam size measurements are needed to determine the three independent Twiss parameters. The use of three independent size measurement locations is easier, but it is possible to use a single beam size measurement station with different upstream focusing conditions. If only a single profile station is available, the upstream focusing should be varied to span a waist at the profile station to ease the fitting of the model.

Figure 4 shows an example of transverse matching. The dots represent measured beam sizes from wire profile devices and the lines show results from an envelope model. The blue lines represent the vertical beam size, and the red lines the horizontal beam size. The upstream Twiss parameters of the beam (at the point marked 'Initial model point' in Fig. 4(a)) were adjusted so that the model beam sizes best matched the measured sizes at the profile measurement stations. Note that in this case the problem is overconstrained, with five profile measurements and three Twiss variables per plane, so the model does not exactly match the measurements. The transport line section shown in Fig. 4(a) includes the end of the superconducting linac (from 150 to 171 m) and the start of a transport line (HEBT) from 171 m to 220 m. The superconducting linac portion of the beam line contains matching quadrupoles, and the HEBT portion is designed to have a FODO lattice structure. The initially measured beam sizes indicate that the beam in the HEBT is not well matched (the FODO lattice structure has a regular periodic beam size variation). Using the envelope model, the quadrupole strengths in the 'matching quadrupoles' region shown in Fig. 4(a) were varied to produce the design Twiss parameters $\alpha$ and $\beta$ at the start of the HEBT. In this step, it is important to include any magnet limits in the solution for the quadrupole settings. Sometimes additional magnets must be employed in the matching process if some quadrupoles are being run near power supply limits. Figure 4(b) shows the measured beam sizes when the new strengths were applied. Now the proper FODO structure is apparent in the HEBT portion of the beam line, as seen by comparing the post-matching results in Fig. 4(b) with Fig. 4(c), which displays the design beam sizes at the beginning of the HEBT.

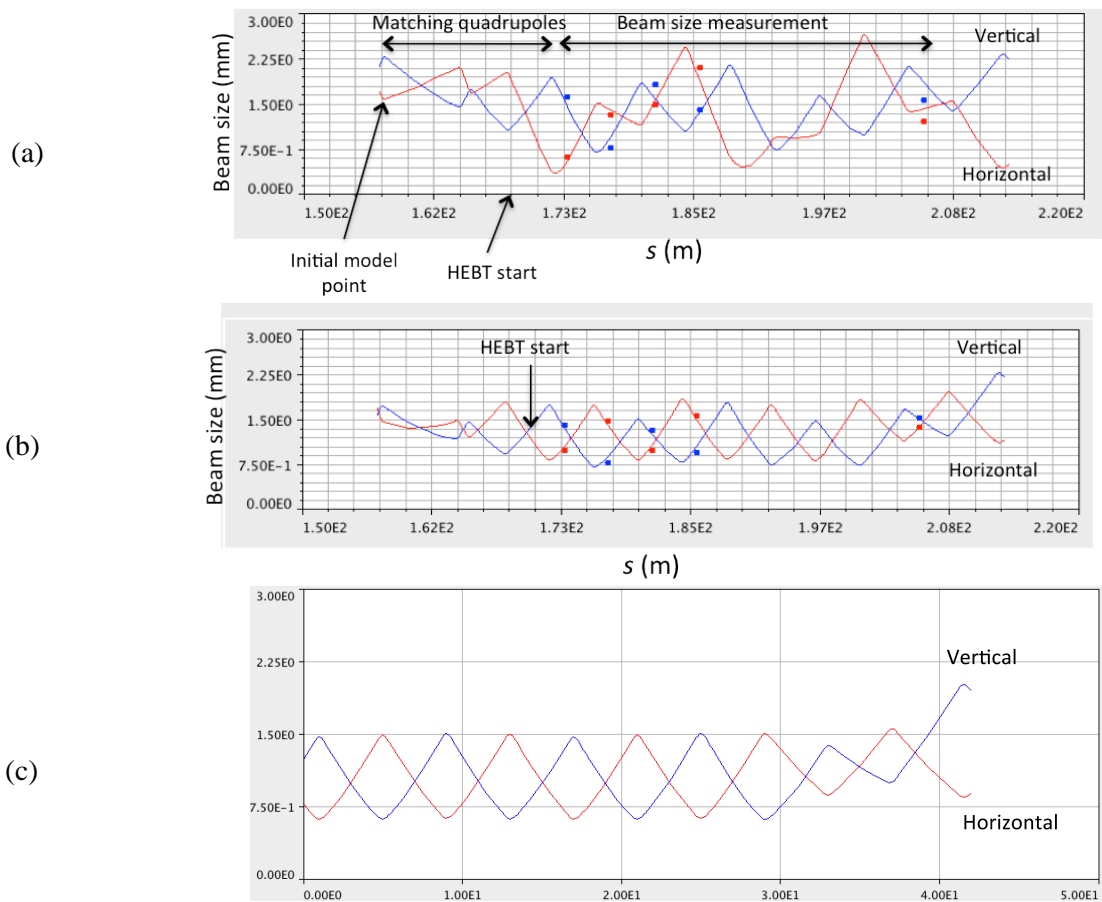

**Fig. 4:** Beam size along the beginning of the SNS HEBT transport line. The blue lines represent the vertical size and the red lines the horizontal size. The dots show measured values, and the lines show envelope model results. (a) Initial Twiss solution, (b) after applying quadrupole corrections, (c) the design lattice.

*3.3.3   Beam response methods*

The advantage of the model-based matching approach described above is that it is faster, as there is a minimal need for beam size measurements. However, it does require a machine-configured model, which may not available. If there is not a machine-configured model available, a beam response technique is sometimes possible. To generate the response matrix, beam sizes are measured while each matching quadrupole is varied independently. From this data ensemble, a response matrix can be constructed and used to provide quadrupole settings that produce the desired beam sizes at the measurement locations. For example, in a FODO structure, with profile measurements arranged to be adjacent to the quadrupoles, equal beam sizes should be expected at each profile station. When one is simply solving for the quadrupole strengths required to produce equal beam sizes at the measurement stations, a beam model is not needed. This approach is used in Refs. [15, 16].

## 4     Longitudinal beam set-up

### 4.1    Longitudinal RF set-up in linacs

In order to properly accelerate a beam in a linac, the phase of the RF field must be properly synchronized with the arrival of the beam at the entrance to the accelerating structure (referred to as a cavity here). For most copper accelerating structures, the amplitude must also be set to a prescribed level. Typical requirements for the RF set-up are a few degrees of RF phase, and about 1% in amplitude. Accurate setting of the accelerating structure is needed to minimize the loss of beam from the accelerating acceptance.

The primary goal is to determine (1) the calibration of the RF drive and the cavity voltage (setting the amplitude) and (2) the timing offset of the RF drive phase with respect to the arrival of the beam (setting the phase). The phase requirement typically translates to a few picoseconds, which requires beam-based solutions. Although a rough calibration of the cavity voltage with the RF drive can be done in advance, beam-based techniques offer resolutions within ~1%. These methods involve varying the RF phase or amplitude or both of a cavity, observing the effect on the beam downstream, and comparing the observed behaviour of the beam with a model prediction.

*4.1.1   Energy degrader approach*

An early approach to the RF set-up was the 'energy degrader' approach [17, 18], in which an intercepting material (degrader) of known thickness is inserted into the beam downstream of a cavity, followed by a charge-measuring device (e.g. a Faraday cup), as shown schematically in Fig. 5. The degrader thickness is chosen to stop an incoming beam with an energy slightly below the design output of the cavity. Thus, only when the output beam is close to being fully accelerated is charge detected. The cavity phase is scanned, and the width of an 'acceptance' can be determined by examining the width of the detected beam signal in the Faraday cup. This process is repeated for several amplitudes, as indicated in Fig. 6(a). The width of the acceptance is plotted versus the RF amplitude, as shown in Fig. 6(b), and the calibration between the RF drive and the actual cavity voltage can be determined by matching this curve with the expected trend based on model predictions for the given degrader thickness. The cavity phase setting is determined relative to the acceptance boundaries at the nominal amplitude setting, again by comparison with model-predicted expectations for the given degrader thickness. Finally, we note that the bunch width of the beam can also be determined from the width of the rise of the curves in the phase scans shown in Fig. 6(a). A drawback of this method is the need for an intercepting device. Also, at higher beam energies (above about 100 MeV for protons), the stopping distance becomes large and this approach becomes impractical.

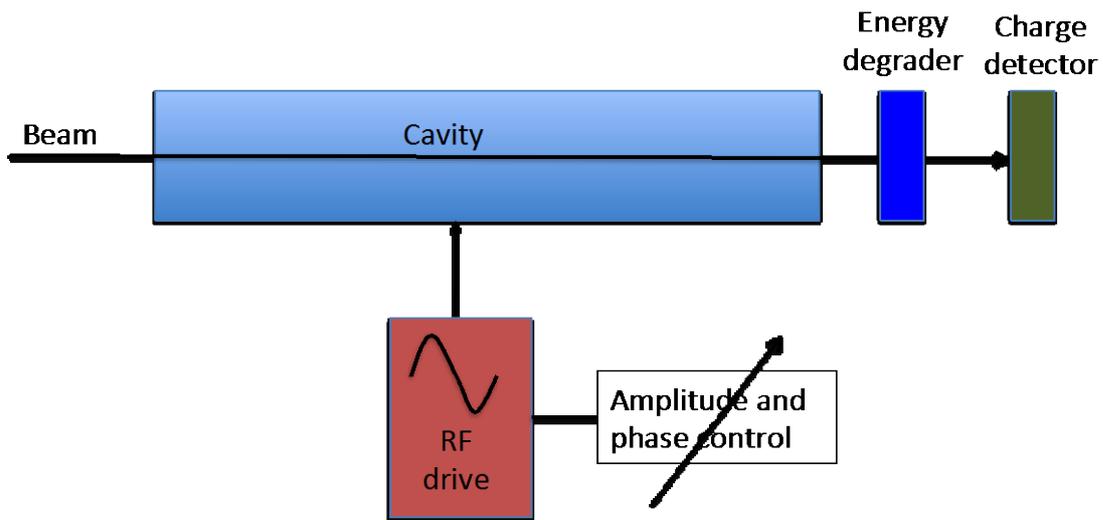

**Fig. 5.** Schematic illustration of the experimental set-up for the energy degrader method

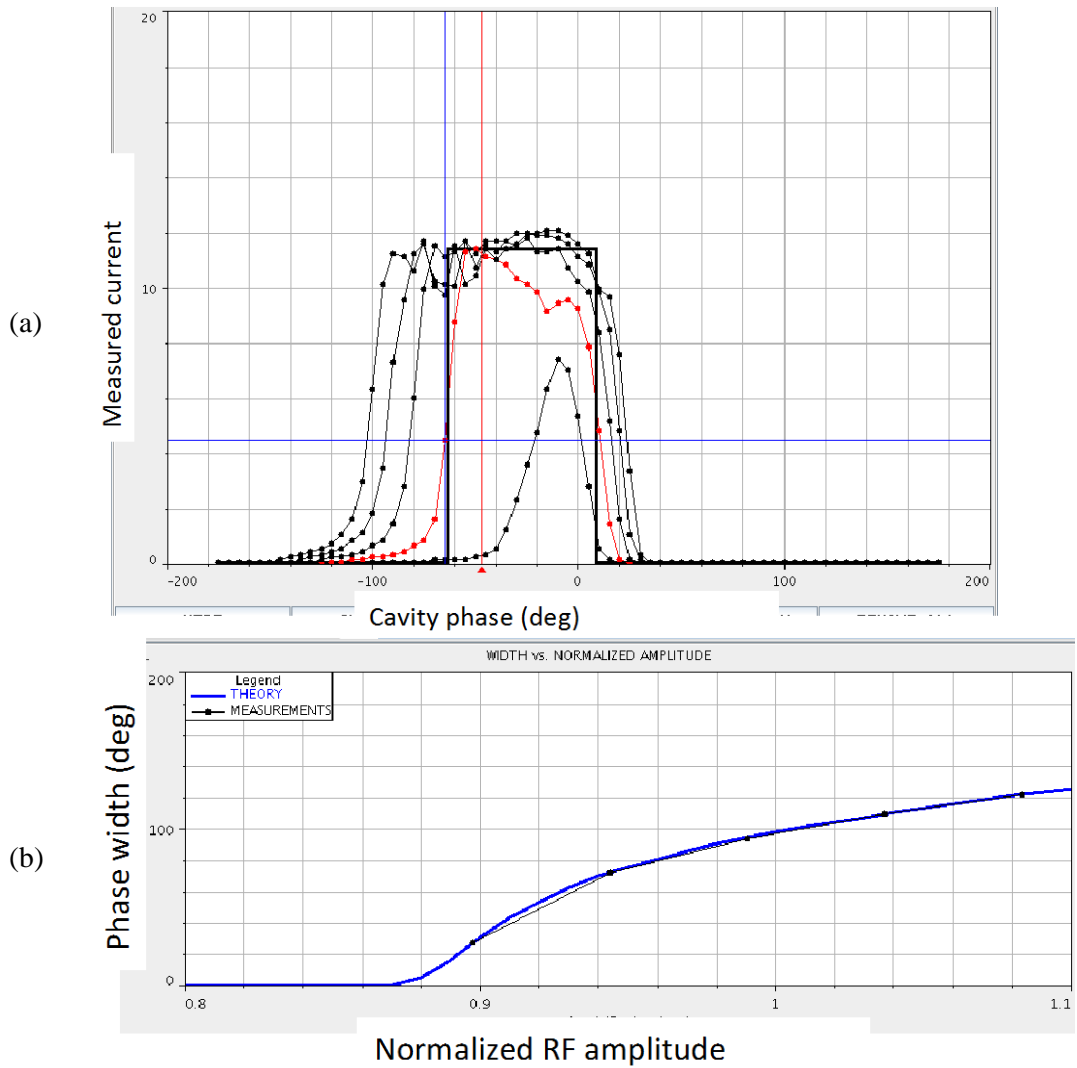

**Fig. 6:** Example of the application of the energy degrader method to a drift tube linac tank in the SNS linac. (a) Measured beam downstream of the degrader vs. RF phase for several amplitudes; (b) width of the detected acceptance windows in the scans vs. the amplitude setting of the RF drive.

*4.1.2  Time-of-flight methods*

A more common approach to setting the RF phases and amplitudes in a linac is a technique in which changes in the beam arrival time downstream of the accelerating structure are measured over a range of RF amplitude and phase settings, and the results are compared with model predictions. By adjusting the model RF settings to best match the observed beam behaviour, one can calibrate the phase and amplitude of the RF hardware. To reduce the sensitivity of the measurements to uncertainties, such as those in the distances between the cavity and the beam pickups, difference techniques are often employed. That is, changes in the Time of Flight (TOF) between two detectors are compared rather than the comparing changes in the beam arrival time at a single detector. The basic concept is shown schematically in Fig. 7. This class of RF set-up is referred to as the TOF method, and includes the Delta-T and signature-matching techniques. All of these techniques require accurate relative-phase measurements between the two beam bunch detectors for varying RF drive conditions. These TOF measurements are typically performed using dedicated diagnostics, for example Fast Current Transformers (FCTs) in the J-PARC linac [19] and specialized treatment of the BPM strip-line signals in the SNS linac [20, 21]. The typical accuracy of the measured changes in the beam TOF is ~1° of RF phase.

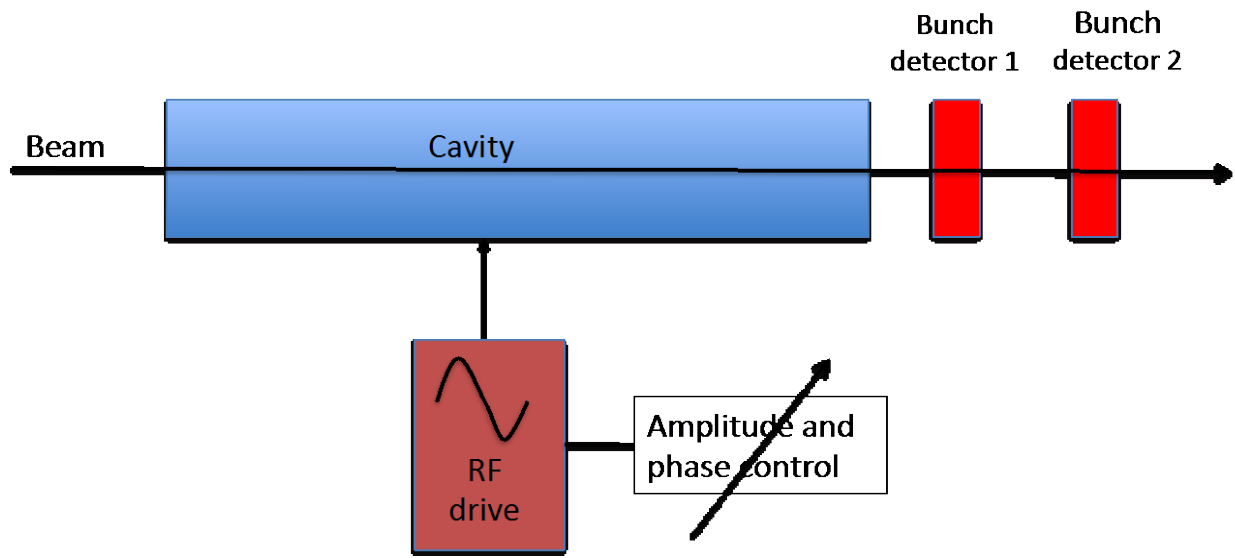

**Fig. 7:** Schematic set-up for beam-based time-of-flight measurements.

In all these approaches, there are three unknowns in the model, which are used to match the observed changes in the TOF with the predictions: (1) the RF amplitude calibration, (2) the RF phase offset, and (3) the energy of the incoming beam in the cavity being tuned. The beam models are typically simple: a beam bunch is treated as a single particle, and the Panofsky equation is solved for the synchronous particle throughout the gaps in the accelerating structure and the subsequent drifts through the TOF detectors.

A concern in the application of these schemes is maintaining the integrity of the bunch as it drifts between the RF structure and the phase detectors. Some pitfalls are beam debunching for low-energy beams (less than a few MeV), which can limit the range of useful RF phase variation. Also, any intervening RF structures between the cavity being tuned and the detectors must be turned off. In the case of superconducting cavities, excitation of intervening cavities by the drifting beam itself can impact on the measured arrival time, and care must be taken to use a beam of low enough intensity or to detune the cavities in some way so that they do not affect the drifting beam.

*4.1.2.1 Delta-T method*

The earliest of these TOF approaches was the Delta-T method [22–25], pioneered at Los Alamos National Laboratory. In this procedure, the predicted variations in arrival time at two downstream detectors B and C are calculated a priori, as difference values:

$$\Delta t_B = (t_{B\,off} - t_{B\,on}) - (t_{B\,off,\,design} - t_{B\,on,\,design}),$$

$$\Delta t_C = (t_{C\,off} - t_{C\,on}) - (t_{C\,off,\,design} - t_{C\,on,\,design}).$$

The subscripts 'off' and 'on' refer to the RF being on and off. The first term on the right-hand side is evaluated for small variations from the design beam energy, the design RF amplitude, and the design RF phase. The second term on the right-hand side is evaluated for the design conditions. An example of such a calculation is shown in Fig. 8. In this figure, there are three curve clusters, each cluster representing different incoming beam energies (nominal and a range of ±50 keV). Within a cluster, each curve represents a different RF amplitude input (±5%, with 1% increments). Along each curve, the RF phase is varied, with each dot representing a 2.5° step.

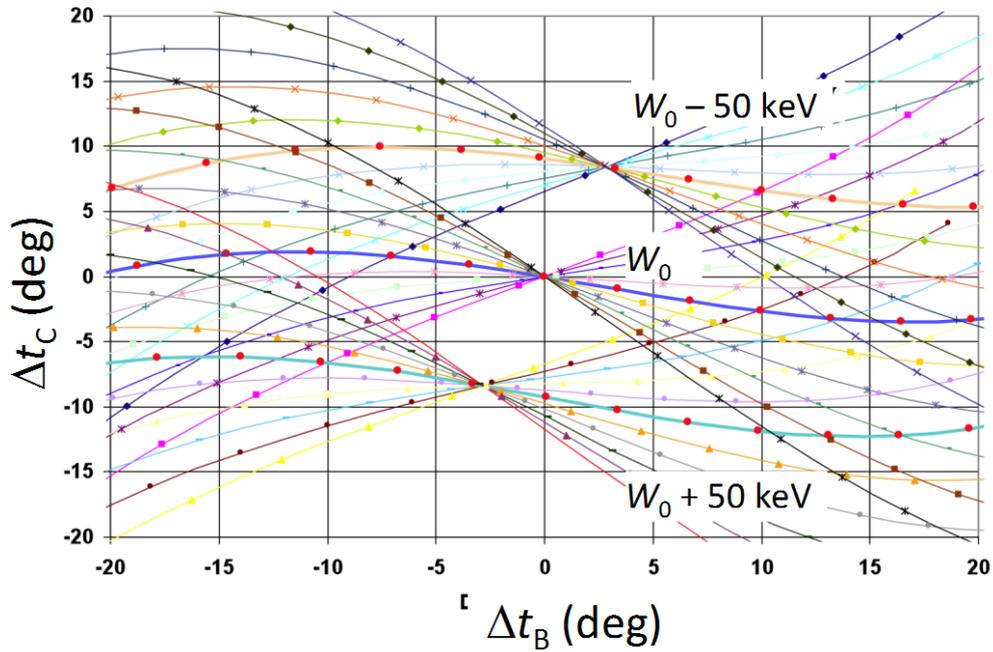

**Fig. 8:** Arrival time difference terms calculated for use in the Delta-T procedure for the SNS Coupled Cavity Linac module 1, with a design input energy of 86 MeV. Note that the time differences here are expressed in units of the detector frequency of 402.5 MHz.

Next, a measured curve of $\Delta t_C$ versus $\Delta t_B$ is obtained by scanning the RF phase and measuring arrival times with detectors B and C, with the RF turned on and turned off. An example measurement is shown in Fig. 9, including a linear fit of the measured values. For small deviations of the incoming beam energy, RF amplitude, and RF phase from their design values, the responses of $\Delta t_B$ and $\Delta t_C$ are linear in these changes. With this assumption, linear algebra techniques can be applied to determine where the measured relation between $\Delta t_B$ and $\Delta t_C$ best lies with respect to the calculated curves, which will have been tabulated. Knowing the RF amplitude and phase offsets, it is straightforward to determine how to change the RF phase and amplitude to reach the design values. The rising straight line in Fig. 9 passes through the locus of points defined by the RF design, but with energy deviations (the centres of the clusters in Fig. 8). The blue line is a linear fit to the measured differences. The intersection of these two lines defines the deviation the beam energy from the design.

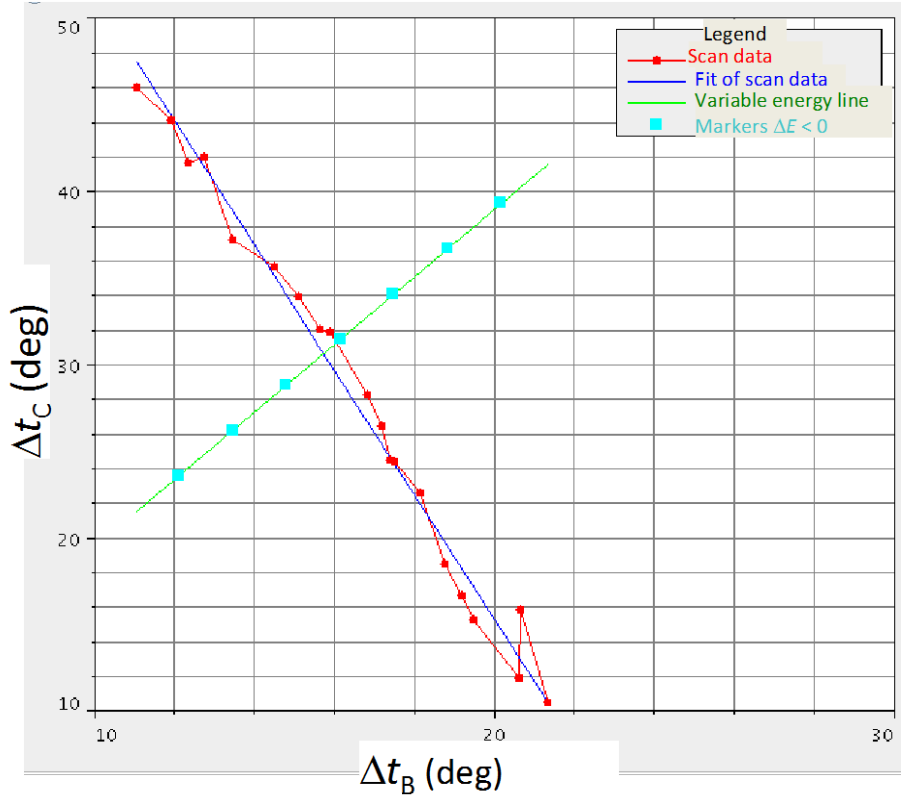

**Fig. 9:** Example of a Delta-T scan of beam phase detector differences for a coupled cavity module in the SNS linac (curve) and a linear fit (descending line).

The processing time for this method is extremely fast, which was an important consideration when the Delta-T method was first deployed, years ago. However, a limitation of this method is the assumption of a linear response of the beam arrival times at detectors B and C to changes in the RF amplitudes and phases. Even in Fig. 8, it is evident that for deviations beyond ±5° or ±10° the response is no longer linear. The method works well when the RF settings are close to the design values, but convergence can be difficult if the initial attempt is far from the design set-up.

*4.1.2.2 Signature matching*

More recently, a generalized approach to the problem of the cavity phase and amplitude has been developed, which is often referred to as a signature-matching scheme. The concept is simple: perform scans across a broad range of RF phase settings, amplitudes, or both, measure the variation in the TOF between two downstream detectors B and C, and use a model to match the measured TOF variation. The method was first demonstrated at the Fermi National Accelerator Laboratory [26], and has subsequently been used at the SNS [27] and J-PARC [28, 29]. This technique requires more computational resources than the Delta-T method, but with modern computers this is not a problem. No linearization of the response of the beam to RF changes is assumed, so this approach is more general and can be applied over wider ranges of RF conditions.

The usual approach is to measure the TOF variation between detectors B and C over a range of RF phases and amplitudes. As in the Delta-T method, measurements are taken with the RF off when possible, and the difference in TOF with the RF on and with it off is tracked:

$$\Delta_{\text{TOF}} = (t_C - t_{C\text{ off}}) - (t_B - t_{B\text{ off}}),$$

where the $t$'s are the measured arrival times with the RF on, and the subscripts 'off' represent the measured arrival times with the RF off. Parametric scans of this quantity at different RF phase and amplitude settings are usually done, producing sets of nonlinear TOF responses. As in the Delta-T method, the model-predicted variations in this quantity are compared with the measured curves. Using optimization methods, the input beam energy, the RF amplitude calibration, and the RF phase offset with respect to the beam are varied to achieve the best match between the predicted and measured curves.

Examples of the application of this technique are shown in Fig. 10. The solid curves show the measured values of the TOF differences, and the dots represent model calculations after the solution was found. The case shown in Fig. 10(a) is for a Drift Tube Linac (DTL) tank at the SNS, with an input energy of about 40 MeV, a longitudinal phase advance of about $2\pi$, 28 accelerating cells, and an energy gain of 17 MeV. RF scans were performed over about 40°, which is considerably more than the range used in the Delta-T method, and the two curves in Fig. 10(a) represent RF amplitude settings which differ by 3.5%. Often, at low energies, the phase scan range is limited by significant beam debunching at incorrect RF settings, which makes arrival time detection problematic. There is a strong nonlinear dependence of $\Delta_{\mathrm{TOF}}$ on the input RF amplitude and phase, and only when the model has the proper calibration and the proper input beam energy do the model and measured values agree. For structures such as this with many accelerating cells, a large change in the beam $\beta$, and a fairly large phase advance, the dependence of $\Delta_{\mathrm{TOF}}$ on the RF settings is quite strong and unique—hence the name 'signature matching' for this method.

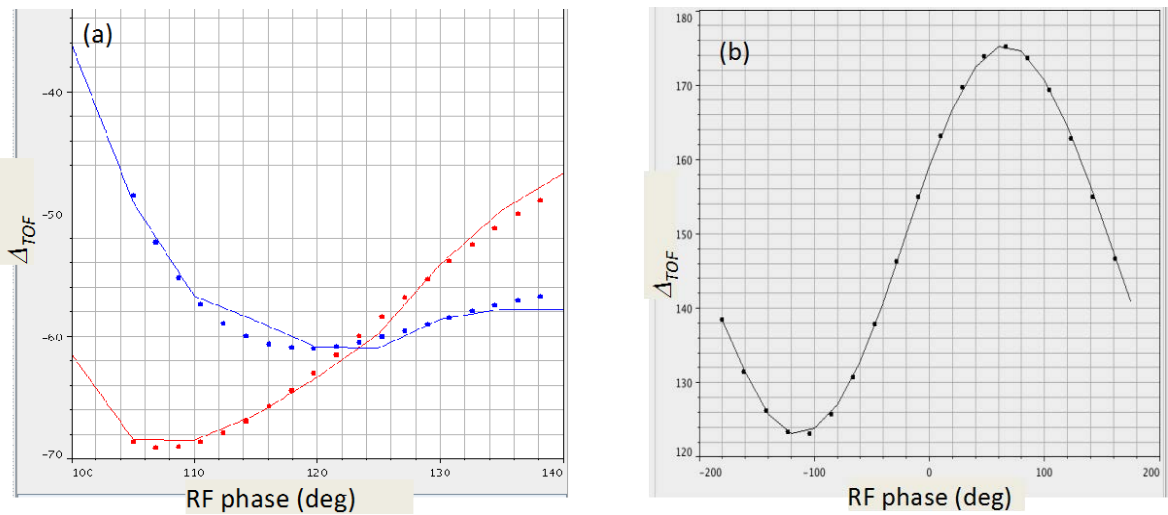

Fig. 10: Examples of the phase scan signature-matching technique for (a) the fourth drift tube linac tank at the SNS, and (b) a typical six-cell superconducting cavity at the SNS.

At high enough beam energies and for short enough accelerating structures, the beam remains bunched well enough for detection over a full 360° variation of the RF phase. An example of this is shown in Fig. 10(b), which is for a superconducting cavity with six cells, an energy gain of about 10 MeV, and a beam energy of about 350 MeV. In this case there is a small change in the beam $\beta$, and a small longitudinal phase advance through the cavity. The resultant variation of $\Delta_{\mathrm{TOF}}$ is nearly sinusoidal, as would be expected from an ideal RF gap kick. The determination of the RF phase and amplitude calibrations and of the incoming beam energy is trivial in this case.

*4.1.2.3 RF shaking*

All of the above techniques are used to determine the proper RF set-up for a single accelerating structure. Often it is useful to check how well a group of accelerating structures is tuned collectively. A simple technique that is useful for this purpose is a beam-based difference technique, analogous to the transverse-orbit-difference technique described in Section 3.2, but in this case in longitudinal space. Typically, a perturbation is applied to the RF drive in an upstream accelerating structure, and the change in beam arrival time is observed throughout a downstream set of accelerating structures. For example, the phase setting of the upstream RF cavity may be changed by 5°. The change in the arrival times of the beam in the downstream phase detectors in the range of cavities being examined is predicted with a model and compared with the measured differences in arrival time. This technique is a useful way of identifying an incorrectly tuned cavity, by observing where the deviation between the measured and predicted differences is initiated. The difference quantity at each detector $i$ measured is

$$\Delta t_{\text{shake } i} = t_{\text{initial } i} - t_{\text{pert } i},$$

where the first term is the beam arrival time at detector $i$ before the RF perturbation and the second term is the beam arrival time after the perturbation. Examples of this procedure are shown in Fig. 11. The first example (Fig. 11(a)) is for the SNS warm linac (a drift tube and coupled cavity structures) and includes an incorrectly set cavity, evidenced by the deviation between the model-predicted change and the observed change. The second example (Fig. 11(b)) shows a case related to the SNS superconducting linac, in which the cavities have properly set RF structures, evidenced by the agreement between the predicted and observed changes.

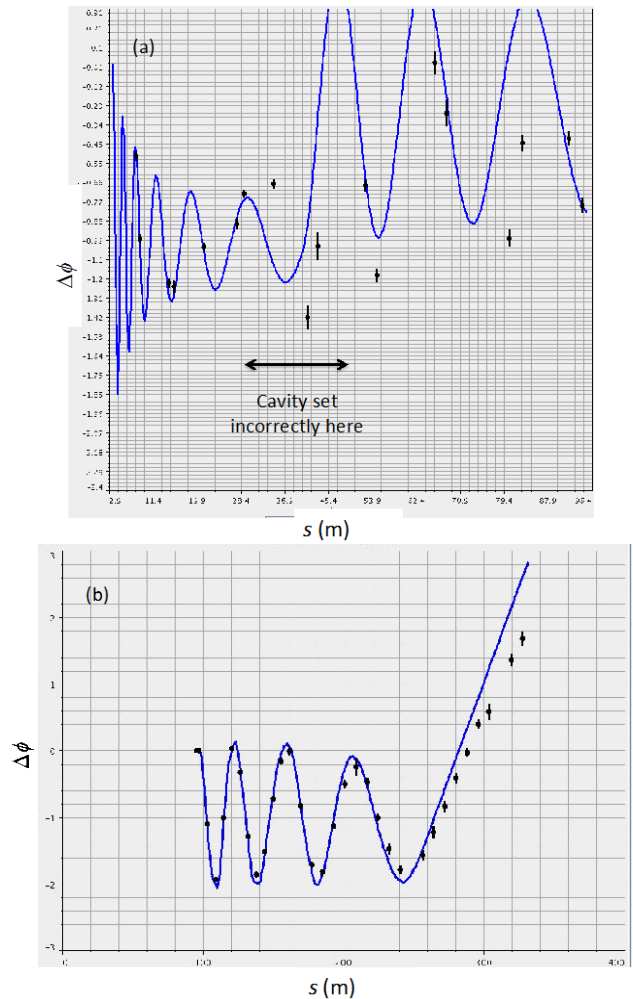

**Fig. 11:** The 'RF shaker' difference technique applied to SNS warm linac structures. (a) A case with an incorrectly set DTL tank; (b) a case of application to the SNS superconducting linac with cavities set correctly.

## 5    Overall commissioning strategies

The above sections have described the preparation for beam commissioning and some fundamental commissioning techniques. There are also some considerations about the commissioning strategy with respect to organizing commissioning schedules. Generally, it is quite difficult to turn on the equipment for a large accelerator complex, so the equipment is kept switched on for long periods. Given that the equipment is on, there is a natural tendency to use it as effectively as possible. At a large institution (e.g. [5]), there are sufficient personnel resources to run it 24 hours a day, seven days a week ('24/7'), with adequate support for the requisite subsystems (e.g. the physics, instrumentation, control, RF, and mechanical subsystems).

At smaller institutions, it is not possible to schedule complete coverage across many groups for extended periods. There are several different ways to handle this situation. At the SNS, the strategy was to have 24/7 coverage of beam physicists, and call in experts as needed [30]. This approach maximizes the utilization of potential beam time, but there is often downtime associated with the time needed for experts to come in when needed. Another approach was taken by J-PARC [30], using 12 h commissioning shifts, but with a broader contingent of experts available. This method allows issues to be addressed faster, but does not fully utilize the potential beam time (note that the machine hardware is still left on all the time). Commissioning has been successfully demonstrated using both approaches.

Another important overall commissioning strategy is to begin commissioning beam line as early as possible. Beam commissioning is the first opportunity for systems to work together in an integrated fashion, and no matter how careful the planning is, there are always surprises. Beam commissioning is the first real test of the ability of systems such as controls, RF systems, timing systems, machine protection systems, and diagnostics to work together, and system interface issues are often identified only when a beam is present. Staging commissioning in such a way that upstream accelerator components can be tested early allows systems (and interfaces) to be shaken out before they are fully deployed throughout the accelerator. Problems identified in a short early commissioning period can be addressed during subsequent downstream equipment installation, and greatly facilitate longer-term progress.

Finally, the most important advice about beam commissioning is to enjoy it. Although it may seem troublesome at the time, it passes quickly and is one of the most exciting periods in the lifetime of an accelerator.

**Acknowledgement**

ORNL is managed by UT-Battelle, LLC, under contract DE-AC05-00OR22725 for the US Department of Energy.